\documentclass[conference]{IEEEtran}
\IEEEoverridecommandlockouts

\usepackage{graphicx}
\usepackage{amsmath}
\usepackage{amssymb}
\usepackage{booktabs}
\usepackage{xcolor}
\usepackage{subcaption}
\usepackage{dblfloatfix}
\usepackage{placeins}
\usepackage{tikz}
\usepackage{xurl}
\usepackage[hidelinks]{hyperref}
\usetikzlibrary{arrows.meta,positioning,calc}

\begin{document}

\title{Computational De Novo Design of Miniprotein Candidate Neutralizers Targeting the IL-23 Cytokine Subunit p19}

\author{\IEEEauthorblockN{Raghav Paladugu\textsuperscript{1}, Fardina Fathmiul Alam\textsuperscript{1,*}}
\IEEEauthorblockA{\textit{\textsuperscript{1}Department of Computer Science, University of Maryland, College Park, MD, USA} \\
rpala06@terpmail.umd.edu, fardina@umd.edu \\
\textsuperscript{*}Corresponding author}
\thanks{Submitted 3 July 2026; accepted 23 September 2026. Accepted for
publication in the Proceedings of the 2026 IEEE International Conference on
Bioinformatics and Biomedicine (BIBM 2026). This is the authors' accepted
version; it is not the IEEE-published version of record.}
\thanks{\copyright~2026 IEEE. Personal use of this material is permitted.
Permission from IEEE must be obtained for all other uses, in any current or
future media, including reprinting/republishing this material for advertising
or promotional purposes, creating new collective works, for resale or
redistribution to servers or lists, or reuse of any copyrighted component of
this work in other works.}}

\maketitle

\begin{abstract}
Interleukin-23 (IL-23) drives T-helper-17 inflammation underlying inflammatory
bowel disease and psoriasis; anti-p19 antibodies are clinically effective but
large, costly, and injectable. De novo miniproteins are an attractive
alternative, yet published designs target the IL-23 \emph{receptor}; the IL-23
\emph{cytokine} itself has not been addressed. A further obstacle is that the
standard confidence score (ipTM) tracks construct length rather than interface
quality. We computationally design miniprotein binders against the
IL-23R-binding face of p19 (PDB 3DUH) and score every design with ipSAE, a
length-robust interface metric derived from AlphaFold-initial-guess predicted
aligned error. A conventional RFdiffusion/ProteinMPNN/ColabFold pipeline yields
no reliable predicted interface across 972 designs (best ipSAE 0.30), and a
scoring artifact---single-sequence ColabFold under-rates binder pLDDT by
$\sim$56 points---explains an earlier misleading ``binders do not fold'' signal.
An interface-optimized hallucination method (BindCraft), scored identically,
yields 42 of 50 designs above ipSAE 0.60 (best 0.804). An AlphaFold-independent
geometric check reproduces the conventional failure: no top-ranked conventional
design reaches the buried area or interface hydrogen-bond count of the weakest
lead. Five all-$\alpha$-helical leads (72--104 residues) collectively engage all
six targeted hotspots; two dock within the intact p40/p19 heterodimer and all
show favorable Rosetta energetics. A p40-alone counter-screen returns ipSAE 0.00
for all five, indicating no spurious p40 interface. To our knowledge these are
the first computationally proposed de novo miniprotein candidates designed to
occlude the IL-23R-binding face of IL-23 p19. All results require experimental
validation.
\end{abstract}


\section{Introduction}
Interleukin-23 (IL-23) is a heterodimeric pro-inflammatory cytokine built from a
shared p40 subunit and an IL-23-specific p19 subunit, the latter a four-helix
bundle~\cite{lupardus2008il23}. Through its receptor it expands a pathogenic
T-helper-17 population that sustains chronic inflammation~\cite{langrish2005il23},
making it among the most validated targets in inflammatory bowel disease (IBD)
and psoriasis: anti-p19 antibodies such as risankizumab and guselkumab are
approved and effective~\cite{dhaens2022risankizumab}. These drugs nonetheless
inherit the antibody format's liabilities, namely high manufacturing cost,
parenteral administration and cold-chain storage, which limit access and comfort.

De novo designed miniproteins are an attractive alternative. At roughly a tenth
the size of an antibody, they can be made hyperstable and resistant to heat, acid
and proteolysis, enabling cheaper production and even oral
delivery~\cite{cao2022minibinder}. This is not merely aspirational for the IL-23
axis: designed miniproteins inhibiting IL-23R and IL-17 recently reached
antibody-like potency, and an orally delivered IL-23R minibinder outperformed a
clinical anti-IL-23 antibody in mouse colitis~\cite{berger2024th17}. That work,
however, targets the receptor and IL-17A. The complementary and arguably more
direct target, the IL-23 cytokine itself neutralized at its IL-23R-binding face
in the manner of an anti-p19 antibody, has not to our knowledge been addressed by
de novo miniprotein design. Targeting p19 rather than p40 also matters
biologically, since p40 is shared across cytokines and a p40-directed agent risks
broader off-target immunomodulation, whereas p19 is unique to
IL-23~\cite{lupardus2008il23,langrish2005il23}. Closing this gap is the first
goal of this paper.

Designing a candidate binder is only half of the problem; the other half is
deciding whether a design is actually likely to bind. Modern pipelines read off
AlphaFold-Multimer's interface ipTM~\cite{evans2022multimer}, which is computed
over whole chains and therefore drifts with construct length rather than
interface quality~\cite{dunbrack2025ipsae,varga2025actifptm}, making it a
treacherous filter for campaigns spanning many binder lengths. The ipSAE metric
restricts the template-modeling expression to interface residue pairs predicted
with low aligned error, so the score reflects the interface and is far less
length-sensitive~\cite{dunbrack2025ipsae}. We adopt ipSAE as our primary success
criterion, interpret it against the published AlphaFold-DB confidence
bands~\cite{afdb} with 0.60 as the reliable-interface threshold and 0.70 as the
confident band, and apply the same scorer to every design from every pipeline so
the comparisons are not confounded by differences in evaluation.

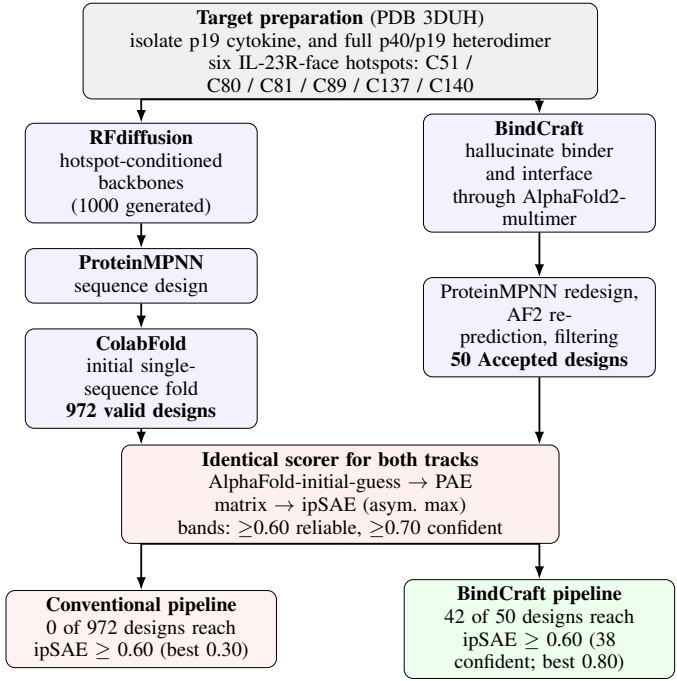
\begin{figure}[!tbp]
  \centering
  \resizebox{\columnwidth}{!}{%
  \begin{tikzpicture}[
    font=\small,
    box/.style={draw, rounded corners, align=center, minimum height=8mm,
                inner sep=3pt, fill=blue!5, text width=34mm},
    score/.style={draw, rounded corners, align=center, minimum height=8mm,
                inner sep=3pt, fill=red!6, text width=66mm},
    outbox/.style={draw, rounded corners, align=center, minimum height=8mm,
                inner sep=3pt, text width=40mm},
    ar/.style={-{Latex[length=2mm]}, thick}]
    \node[box, fill=gray!12, text width=78mm] (tgt) at (0,0)
      {\textbf{Target preparation} (PDB 3DUH)\\isolate p19 cytokine, and full p40/p19 heterodimer\\six IL-23R-face hotspots: C51 / C80 / C81 / C89 / C137 / C140};
    \node[box] (rf)   at (-3.1,-1.9) {\textbf{RFdiffusion}\\hotspot-conditioned backbones\\(1000 generated)};
    \node[box] (mpnn) at (-3.1,-3.5) {\textbf{ProteinMPNN}\\sequence design};
    \node[box] (cf)   at (-3.1,-5.1) {\textbf{ColabFold}\\initial single-sequence fold\\\textbf{972 valid designs}};
    \node[box] (bc)   at (3.1,-1.9) {\textbf{BindCraft}\\hallucinate binder and interface\\through AlphaFold2-multimer};
    \node[box] (bc2)  at (3.1,-4.3) {ProteinMPNN redesign,\\AF2 re-prediction, filtering\\\textbf{50 Accepted designs}};
    \node[score] (ig) at (0,-6.9)
      {\textbf{Identical scorer for both tracks}\\AlphaFold-initial-guess $\rightarrow$ PAE matrix $\rightarrow$ ipSAE (asym.\ max)\\bands: $\geq$0.60 reliable, $\geq$0.70 confident};
    \node[outbox, fill=red!5]   (r1) at (-3.1,-9.0) {\textbf{Conventional pipeline}\\0 of 972 designs reach\\ipSAE $\geq$ 0.60 (best 0.30)};
    \node[outbox, fill=green!7] (r2) at (3.1,-9.0)  {\textbf{BindCraft pipeline}\\42 of 50 designs reach\\ipSAE $\geq$ 0.60 (38 confident; best 0.80)};
    \draw[ar] (tgt.south) -| (rf.north);
    \draw[ar] (tgt.south) -| (bc.north);
    \draw[ar] (rf)   -- (mpnn);
    \draw[ar] (mpnn) -- (cf);
    \draw[ar] (bc)   -- (bc2);
    \draw[ar] (cf.south)  -- (cf.south |- ig.north);
    \draw[ar] (bc2.south) -- (bc2.south |- ig.north);
    \draw[ar] (ig.south) -| (r1.north);
    \draw[ar] (ig.south) -| (r2.north);
  \end{tikzpicture}}
  \caption{Pipeline overview and the central result. Both tracks begin from the same IL-23
p19 target and the same six IL-23R-face hotspots, and every design from both
tracks is judged by one identical interface scorer (AlphaFold-initial-guess PAE
processed into ipSAE, asymmetric maximum), so the comparison reflects the design
method rather than differences in evaluation. Reading the outcome boxes against
the design counts above them shows the conventional pipeline did not fail for
want of designs.}
  \label{fig:pipeline}
\end{figure}

This paper makes three contributions: we characterize a scoring artifact in which
single-sequence ColabFold severely under-rates designed-binder foldability
relative to AlphaFold-initial-guess, producing a spurious ``binders do not fold''
signal; we show under rigorous interface-aware scoring that a standard
RFdiffusion/ProteinMPNN/ColabFold pipeline yields no reliable predicted interface
against the IL-23R face of p19 across 972 designs; and we show that
interface-optimized hallucination, scored identically, yields 42 of 50 designs
above the 0.60 threshold (38 confident), characterizing five leads structurally
and energetically. Figure~\ref{fig:pipeline} summarizes both tracks and the shared scorer.

\section{Related Work}
\textbf{Deep-learning protein design.} Early deep generative and
representation-learning models showed that neural networks can capture protein
tertiary geometry and generate physically realistic structures and distance
maps~\cite{alam2021latent,alam2022datasize}, a line that culminated in
AlphaFold2~\cite{jumper2021alphafold} and
AlphaFold-Multimer~\cite{evans2022multimer}, commonly run through
ColabFold~\cite{mirdita2022colabfold}. For design, RFdiffusion generates
backbones by denoising diffusion~\cite{watson2023rfdiffusion} and ProteinMPNN
assigns sequences likely to fold into them~\cite{dauparas2022proteinmpnn};
together they form the dominant fixed-backbone pipeline, our baseline. Two ideas
are central to \emph{evaluating} designs. AlphaFold-initial-guess scores designed
complexes by seeding the prediction with the designed pose, raising experimental
success roughly tenfold as a filter~\cite{bennett2023initialguess}. BindCraft
instead hallucinates binder and interface together by backpropagating through
AlphaFold2, co-folding target and binder so the interface can relax into place,
reporting 10--100\% experimental success across diverse
targets~\cite{pacesa2025bindcraft}. This contrast, score-after-design versus
design-by-scoring, is what our comparison isolates.

\textbf{Interface confidence metrics.} The default score, ipTM, is computed over
entire chains and drifts with construct length, the wrong behaviour when binders
of many sizes are compared. actifpTM restricts the pTM calculation to flexible,
interface-participating residues~\cite{varga2025actifptm}, and ipSAE derives a
template-modeling-style score from predicted aligned error over only confidently
predicted interchain pairs~\cite{dunbrack2025ipsae}. We adopt ipSAE and validate
our implementation against the reference code on its distributed example.

\textbf{IL-23 biology and miniprotein therapeutics.} The 2.3-\AA{} structure of
IL-23 (3DUH) established the p19/p40 architecture~\cite{lupardus2008il23}.
Designed miniproteins against IL-23R and IL-17A established the modality for the
Th17 axis~\cite{berger2024th17}; we extend it to the IL-23 cytokine.

\section{Methods}
\textbf{A. Target preparation.} IL-23 coordinates were taken from PDB 3DUH
(chain A $=$ p40, chain C $=$ p19; p19 modeled at residues 1--29, 47--94,
100--125, 137--169). We defined two targets: the isolated p19 cytokine (chain C,
136 modeled residues) and the intact p40/p19 heterodimer. Six residues on the
IL-23R-binding face of p19 were selected as design hotspots: C51, C80, C81, C89,
C137, C140 (3DUH chain-C numbering), all in modeled regions.

\textbf{B. Conventional pipeline (baseline).} Backbones were generated with
RFdiffusion conditioned on the six p19 hotspots (C51/C80/C81/C89/C137/C140),
using 50 diffusion timesteps (\texttt{diffuser.T}$=$50) on a constant CA/frame
noise schedule with the main denoiser noise set to zero
(\texttt{noise\_scale\_ca}$=$\texttt{noise\_scale\_frame}$=$0), a hotspot guide
scale of 10 and no auxiliary guiding potential; 1{,}000 backbones were generated
(972 valid after filtering and sequence threading), followed by a
partial-diffusion refinement pass (\texttt{partial\_T}$=$20). Sequences were
assigned with ProteinMPNN (soluble-model weights, interface fixed, sampling
temperature 0.1, zero backbone noise; 20 sequences sampled per backbone, best 2
retained), and complexes were initially folded with ColabFold~1.5.5
(\texttt{alphafold2\_multimer\_v3}, single-sequence MSA, 5 models, 3 recycles).

\textbf{C. Interface-aware re-scoring.} Every design was re-evaluated with
AlphaFold-initial-guess, seeding the binder--p19 complex and recovering
per-residue pLDDT and the full predicted-aligned-error matrix. From the PAE we
computed ipSAE following the reference definition, reporting the maximum over the
two asymmetric chain directions (Dunbrack's Eq.~16 and the headline value emitted
by the reference \texttt{ipsae.py}~\cite{dunbrack2025ipsae}). Each score comes
from a single pose-seeded prediction, not averaged across model seeds, and our
implementation reproduces \texttt{ipsae.py} to full numerical agreement on its
distributed example as a code-correctness check. For interpretation we adopt the
AlphaFold-DB ipSAE bands ($\geq$0.80 very high, 0.70--0.80 confident, 0.60--0.70
possibly correct, $<$0.60 unlikely)~\cite{afdb}, taking 0.60 as the
reliable-interface threshold and reporting the 0.70 count alongside. Because the
maximum is at least the per-direction minimum, applying 0.60 to the maximum is
permissive; this affects neither the five leads, all above 0.70, nor the pipeline
contrast, since the conventional best (0.297) falls below every threshold under
either convention. As a direct check, the two leads whose initial-guess
predictions were retained (l94, l104) give minimum-direction ipSAE of 0.748 and
0.801, both still confident.

\textbf{D. Interface-optimized pipeline (BindCraft).} BindCraft was run against
both the p19-only and full-heterodimer targets with the same six hotspots, using
the \texttt{default\_4stage\_\allowbreak multimer\_\allowbreak
flexible\_\allowbreak hardtarget} protocol: the binder is hallucinated against
the co-folded target by backpropagation through AlphaFold2-multimer, redesigned
with ProteinMPNN, then re-predicted and filtered, with a flexibly modeled
(relaxable) target and stricter hard-target acceptance. We used this variant
because the standard protocol collapsed at the sequence-discretization stage on
this target. The binder-length range was 65--130 residues. The four design stages
used 75 soft, 45 temporary, 5 hard and 15 greedy iterations; the multimer-design
loss weighted pLDDT 0.1, intra-chain PAE 0.4, inter-chain PAE 0.1, intra- and
inter-chain contacts 1.0 each, helicity $-0.3$, ipTM 0.05 and radius of gyration
0.3, with the target sequence removed during design and prediction (side chains
retained) and $\beta$-sheet optimization enabled. AlphaFold2 used 1 recycle
during design and 3 at validation. BindCraft reports per-design Rosetta interface
metrics ($\Delta$G, buried area, shape complementarity, packing, interface
hydrogen bonds, hotspot RMSD, helicity).

\textbf{E. Apples-to-apples comparison.} Every BindCraft Accepted design was
re-scored through the identical AlphaFold-initial-guess $+$ ipSAE pipeline used
for the 972 conventional designs (binder relabeled, p19 $=$ the C-terminal 136
residues of the target); only the design method differs. Two scope notes apply.
First, the comparison is between valid conventional outputs (972 surviving
filtering and sequence threading) and BindCraft Accepted outputs (50 meeting its
internal criteria), so it compares per-output interface quality under a common
scorer rather than raw generation efficiency. Across both targets BindCraft ran
174 trajectories and 1{,}984 ProteinMPNN sequences (136/1{,}635 p19-only, 38/349
full heterodimer), of which 50 were Accepted. To express the contrast on a common
budget we also report yield per attempt: the conventional pipeline retained two
of twenty sequences per backbone, giving $\approx$1{,}944 scored sequences
against BindCraft's 1{,}984. Reliable designs per retained sequence are 0/1{,}944
versus 42/1{,}984 (2.1\%), and per trajectory 0/1{,}000 versus 42/174 (24.1\%),
so the contrast does not rest on comparing filtered against unfiltered output.
Second, because BindCraft optimizes against AlphaFold2 internally, the external
scorer is not fully independent of its objective, whereas it is independent of
the RFdiffusion/ProteinMPNN pipeline; this could modestly favor BindCraft, though
the conventional pipeline produced no design above any tested threshold
(Table~\ref{tab:sens}) and the gap also reproduces under AlphaFold-independent
geometry (Table~\ref{tab:geom}).

\textbf{F. Lead characterization.} For each lead we determined interface contacts
(binder heavy atom within 5\,\AA{} of p19), mapped them to canonical 3DUH chain-C
numbering to identify engaged hotspots, computed secondary structure by DSSP via
PyRosetta~\cite{chaudhury2010pyrosetta,kabsch1983dssp}, and recorded the
BindCraft Rosetta interface metrics.

\textbf{G. AlphaFold-independent interface geometry.} The 20 top-ranked
conventional designs with retained AlphaFold-initial-guess structures were
analysed with the Rosetta \texttt{InterfaceAnalyzerMover} (binder versus target
chain, separated-pose packing), recording buried area, shape complementarity and
interface hydrogen bonds. Because 41 higher-ranked structures were not retained,
this set spans ipSAE 0.011--0.225 and represents the upper range rather than its
strict maximum. We report geometric descriptors only; Rosetta $\Delta$G is
omitted because AlphaFold outputs are not energy-minimized, so unrelaxed
$\Delta$G reflects strain rather than binding and is not comparable to
BindCraft's post-relaxation values.

\textbf{H. Compute.} All design and scoring ran on an HPC cluster with NVIDIA
A100 (40\,GB) GPUs under SLURM.
Binder sequences, scaffold data, and scripts are available in the
\href{https://github.com/rpalad/ieee_bibm_submission_supplementals}
{supplementary GitHub repository}
(\url{https://github.com/rpalad/ieee_bibm_submission_supplementals}).

\section{Results}

\subsection{The ColabFold scoring artifact}
\begin{figure}[!tbp]
  \centering
  \includegraphics[width=0.804\columnwidth]{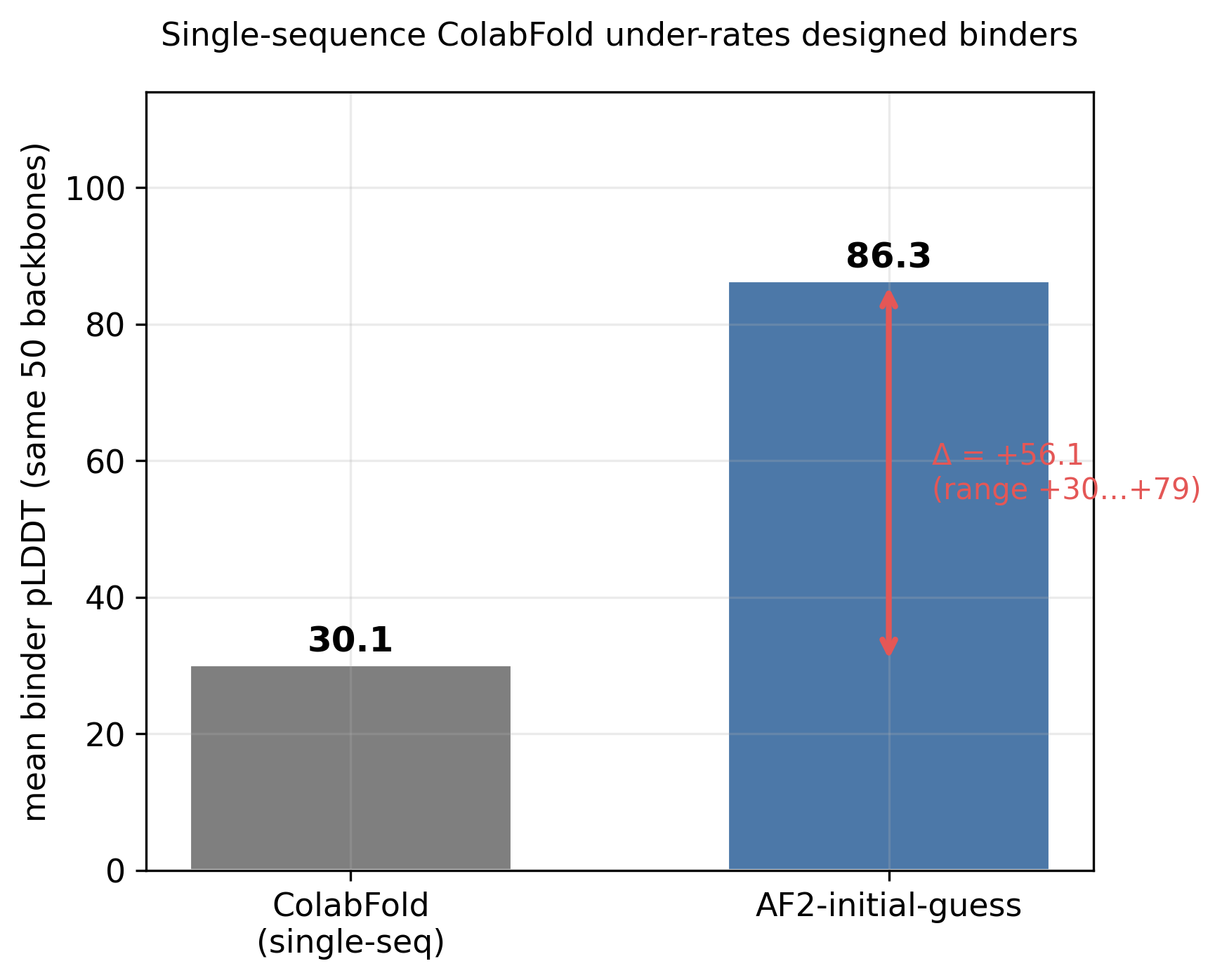}
  \caption{The single-sequence ColabFold scoring artifact. For 50 designs the \emph{same}
backbone is folded two ways: single-sequence ColabFold rates the binders as
essentially disordered (mean pLDDT 30.1) while AlphaFold-initial-guess, seeded
with the designed pose, rates them well-folded (mean 86.3). The gap is a property
of the scorer. Source data: Supplementary~S6.}
  \label{fig:artifact}
\end{figure}

An early reading suggested the designed binders were intrinsically disordered:
single-sequence ColabFold assigned them very low binder pLDDT. Re-folding the
\emph{same} backbones with AlphaFold-initial-guess overturned this: on 50 matched
designs mean binder pLDDT was 30.1 versus 86.3, a mean under-rating of 56 points
(per-design range $+30$ to $+79$; Fig.~\ref{fig:artifact}), and ColabFold's ipTM
was essentially uncorrelated with ipSAE (Spearman $\approx -0.07$). We stress the
direction of this result: single-sequence ColabFold is the unreliable reading we
reject, while pose-seeded AlphaFold-initial-guess with ipSAE is what we adopt
throughout. The ``unfoldable binder'' signal was therefore an artifact of
single-sequence scoring, not evidence of monomer failure; the open question is
the \emph{interface}, not the monomer.

\subsection{Conventional pipeline finds no reliable predicted interface}
\begin{figure}[!tbp]
  \centering
  \vspace{-12pt}
  \includegraphics[width=0.97\columnwidth]{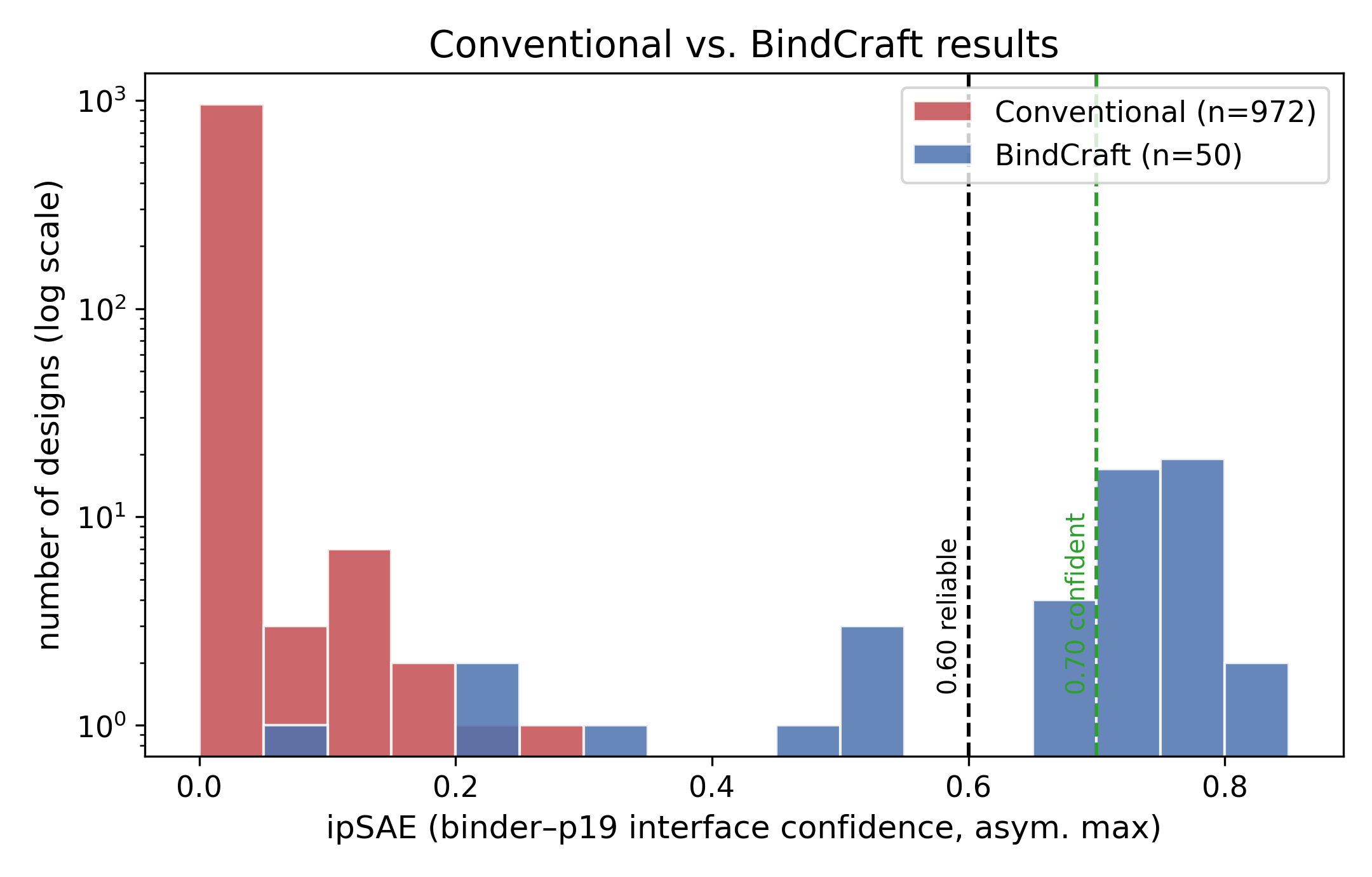}
  \caption{ipSAE distribution for every design from both pipelines, shared axis and
  logarithmic count scale. The conventional distribution is collapsed against zero,
  reaching neither the 0.60 threshold (black dashed) nor the 0.70 band (green
  dashed), whereas BindCraft sits largely above both. Because the conventional
  designs nonetheless have high predicted monomer confidence
  (Fig.~\ref{fig:artifact}), the failure is localized to the interface rather than
  backbone quality. Source data: Supplementary~S5.}
  \label{fig:convdist}
  \vspace{-12pt}
\end{figure}

With foldability resolved, we scored interfaces directly. Across the full
972-design screen no design reached the threshold: the best binder--p19 ipSAE was
0.297, the median 0.005, and 0/972 reached 0.60 (none reached even 0.40), with
the distribution collapsed toward zero (Fig.~\ref{fig:convdist}). A curated
top-50 subset re-scored against both subunits engaged neither, the best p19
interface reaching ipSAE 0.235 and the best off-target p40 interface 0.147. Since
these designs have high AlphaFold-initial-guess binder pLDDT ($\sim$84--92,
median $\sim$89) yet form no reliable interface, the limitation is interface
optimization rather than backbone quality, precisely the failure mode an
interface-aware scorer exposes. The failure is also size-independent: ipSAE shows
no correlation with binder length across all 972 designs (Spearman $-0.05$).

\textbf{An AlphaFold-independent check.} Because both ipSAE and BindCraft's
objective derive from AlphaFold2, the contrast could in principle reflect the
scorer rather than the designs. We therefore re-evaluated the conventional
designs using interface geometry alone, which involves no AlphaFold-derived
quantity (Table~\ref{tab:geom}). None of the 20 top-ranked designs reaches the
buried area or hydrogen-bond count of even the weakest BindCraft lead, 12 of 20
form no interface hydrogen bond, and 0 of 20 meet all three lead-range minima
jointly; the one design with lead-comparable shape complementarity (0.695) buries
only 144\,\AA$^2$, a small packed patch rather than an interface. The failure
thus reproduces under a purely geometric criterion, indicating a property of the
designs rather than an artifact of AlphaFold-based scoring. These values are
computed against the merged p40/p19 target, which offers \emph{more} surface to
bury than the p19-only target used for three leads, so the comparison is
conservative.

\begin{table}[!tbp]
  \centering
  \caption{AlphaFold-independent interface geometry for the 20 top-ranked conventional
designs with retained structures; median (range). No conventional design reaches
the weakest lead on buried area or hydrogen bonds, so the pipeline contrast does
not depend on AlphaFold-derived scoring. Conventional values are scored against
the merged p40/p19 target, which offers more surface to bury, making the
comparison conservative. Metrics are geometric only; $\Delta$G is excluded
(Methods~G).}
  \label{tab:geom}
  \footnotesize
  \setlength{\tabcolsep}{3pt}
  \renewcommand{\arraystretch}{1.1}
  \begin{tabular}{lcc}
    \toprule
           & Conventional      & BindCraft \\
    Metric & ($n{=}20$)        & leads ($n{=}5$) \\
    \midrule
    dSASA (\AA$^2$)   & 1149 (144--1755)  & 2204--3018 \\
    Shape compl.      & 0.44 (0.07--0.70) & 0.62--0.69 \\
    Interface H-bonds & 0 (0--3)          & 3.5--10.5 \\
    Meets all three   & 0/20              & 5/5 \\
    \bottomrule
  \end{tabular}
\end{table}

\subsection{Interface-optimized design clears the reliability threshold}
\begin{figure}[!tbp]
  \centering
  \begin{subfigure}{\columnwidth}
    \centering
    \includegraphics[width=0.770\columnwidth]{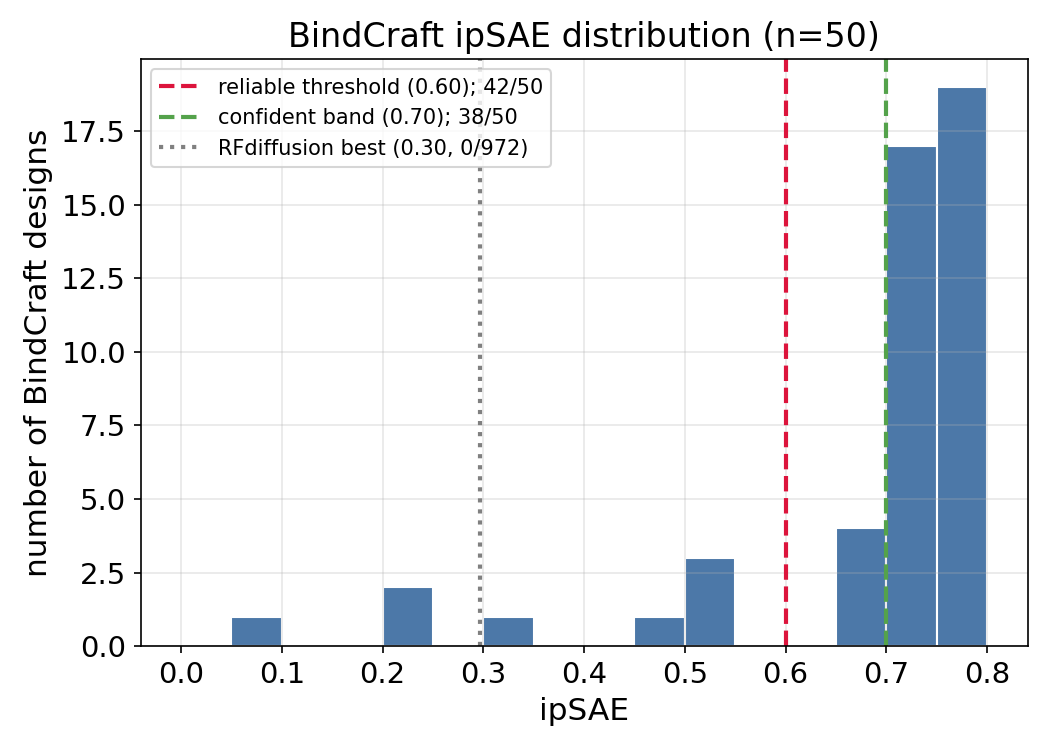}
    \caption{}
    \label{fig:bindcraft-dist}
  \end{subfigure}
  \\[-2pt]
  \begin{subfigure}{\columnwidth}
    \centering
    \includegraphics[width=0.770\columnwidth]{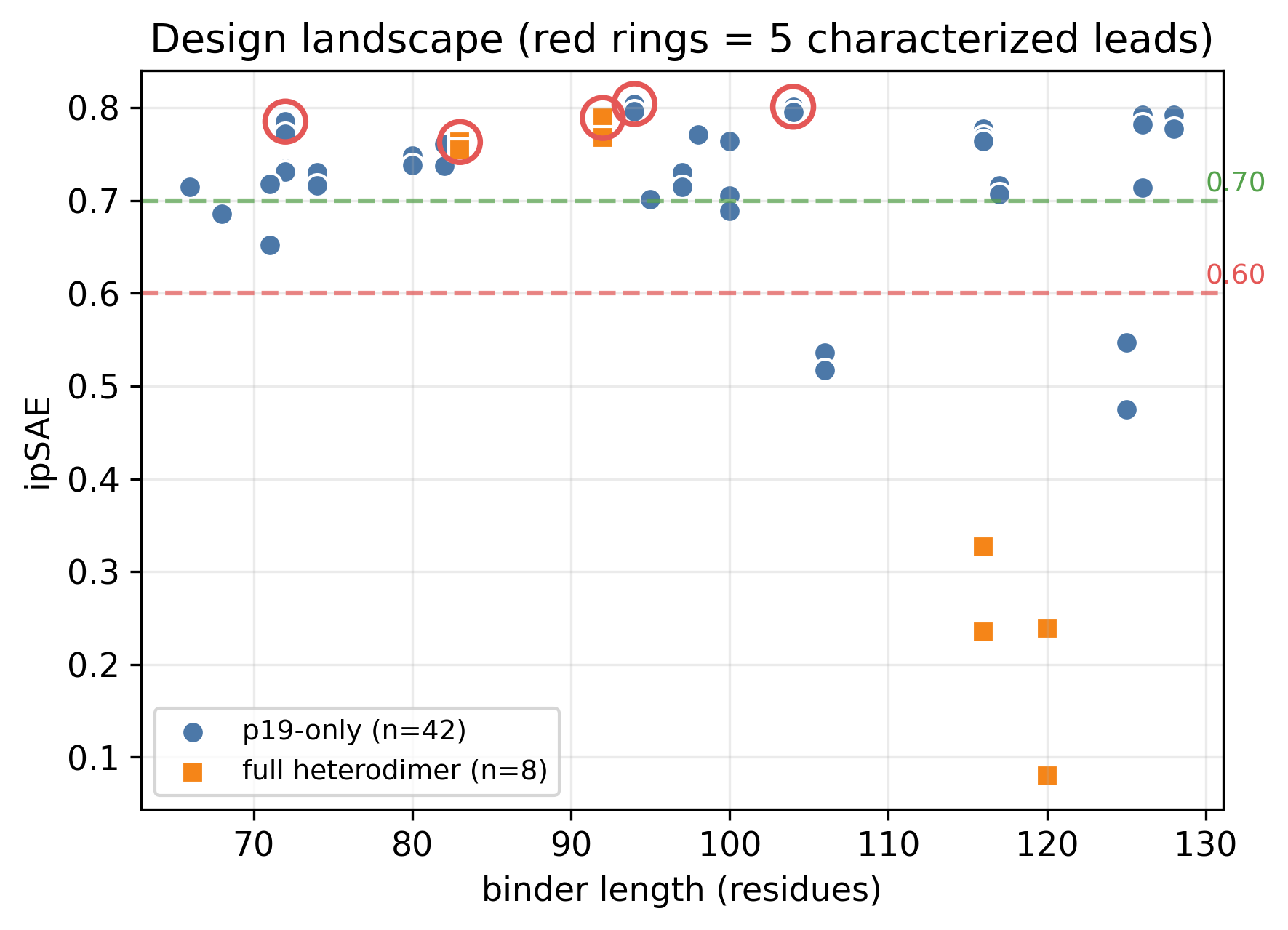}
    \caption{}
    \label{fig:bindcraft-land}
  \end{subfigure}
  \caption{BindCraft outcome under the shared scorer. (a)~Distribution of binder--p19 ipSAE
over the 50 designs; dashed lines mark the 0.60 and 0.70 thresholds, the
conventional best (dotted) sits far left. (b)~ipSAE versus binder length by
regime; rings mark the five leads.}
  \label{fig:bindcraft}
\end{figure}

Re-running the campaign with BindCraft under the identical
AlphaFold-initial-guess $+$ ipSAE pipeline changed the outcome sharply. Of 50
Accepted designs (42 p19-only, 8 full heterodimer), 42 cleared the 0.60
threshold, 38 of them in the $\geq$0.70 confident band and 2 in the $\geq$0.80
band; the best reached ipSAE 0.804 and the minimum interface PAE fell below
1\,\AA{} (Fig.~\ref{fig:bindcraft}a). The reliable designs span 23 independent
seeds (Supplementary~S2), so the result is not one fortunate scaffold. The eight
designs below 0.60 are the four weakest full-heterodimer designs (ipSAE
0.08--0.33) plus four p19-only designs (0.47--0.55). The headline contrast under
identical scoring is 0/972 (best 0.297) versus 42/50: a two-sided Fisher exact
test gives $p \approx 7\times10^{-67}$ and the success rates do not overlap
(Wilson 95\% CI 0.72--0.92 versus 0--0.004). It is also robust to the threshold
(Table~\ref{tab:sens}): at every cutoff from 0.50 to 0.80 the conventional
pipeline yields zero designs while BindCraft yields many.

The design landscape (Fig.~\ref{fig:bindcraft}b) shows reliable designs across
the observed 66--128-residue range, with the p19-only regime almost entirely
above threshold (38/42) and the four full-heterodimer failures forming the
lowest-scoring tail.

\begin{table}[!tbp]
  \centering
  \caption{Threshold-sensitivity of the pipeline contrast: designs at or above each ipSAE
cutoff. The conventional pipeline yields zero at every cutoff, so the contrast
does not depend on the 0.60 choice.}
  \label{tab:sens}
  \small
  \setlength{\tabcolsep}{8pt}
  \renewcommand{\arraystretch}{1.2}
  \begin{tabular}{ccc}
    \toprule
    ipSAE cutoff & Conventional ($n{=}972$) & BindCraft ($n{=}50$) \\
    \midrule
    $\geq$0.50 & 0 & 45 \\
    $\geq$0.60 & 0 & 42 \\
    $\geq$0.70 & 0 & 38 \\
    $\geq$0.80 & 0 & 2  \\
    \bottomrule
  \end{tabular}
\end{table}

\textbf{Scaffold and sequence diversity.} The 42 p19-only designs are not
near-duplicates: they derive from 23 distinct BindCraft backbones spanning 19
binder lengths between 66 and 128 residues (Supplementary~S2). Across the 861
design pairs the median pairwise sequence identity is 27\% (mean 28\%), only
2.2\% of pairs exceed 50\% identity, and single-linkage clustering at 50\%
identity yields 23 clusters, exactly the number of distinct backbones, so the
only highly similar pairs are alternative ProteinMPNN sequences of one scaffold.

\subsection{Five lead candidates cover the targeted epitope}
\begin{figure}[!tbp]
  \centering
  \begin{subfigure}{\columnwidth}
    \centering
    \includegraphics[width=0.770\columnwidth]{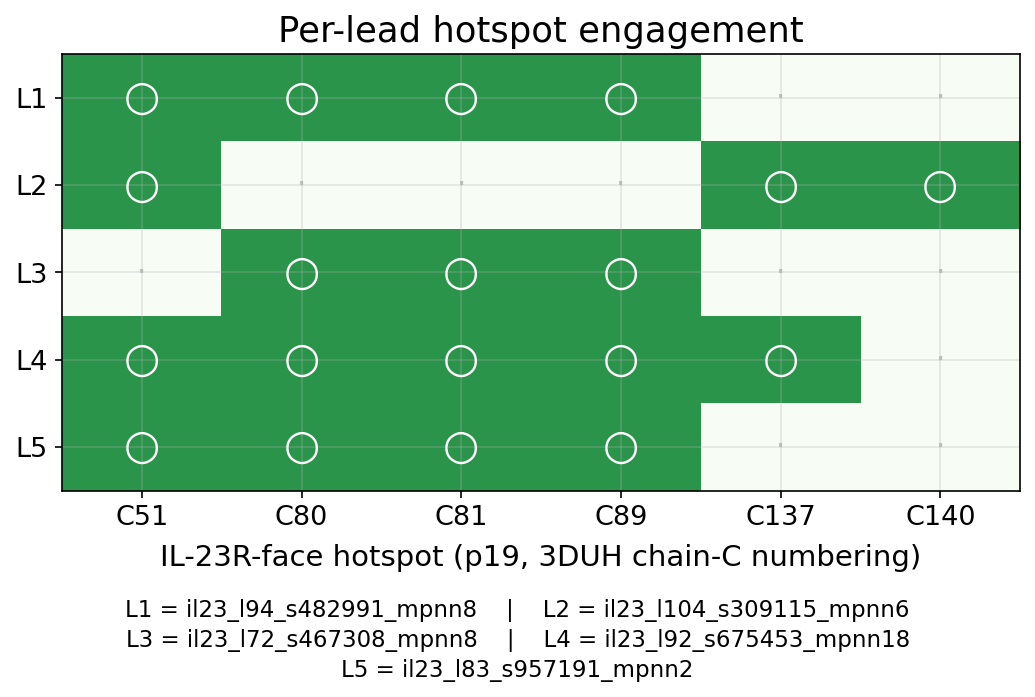}
    \caption{}
    \label{fig:leads-hotspot}
  \end{subfigure}
  \\[-2pt]
  \begin{subfigure}{\columnwidth}
    \centering
    \includegraphics[width=0.653\columnwidth]{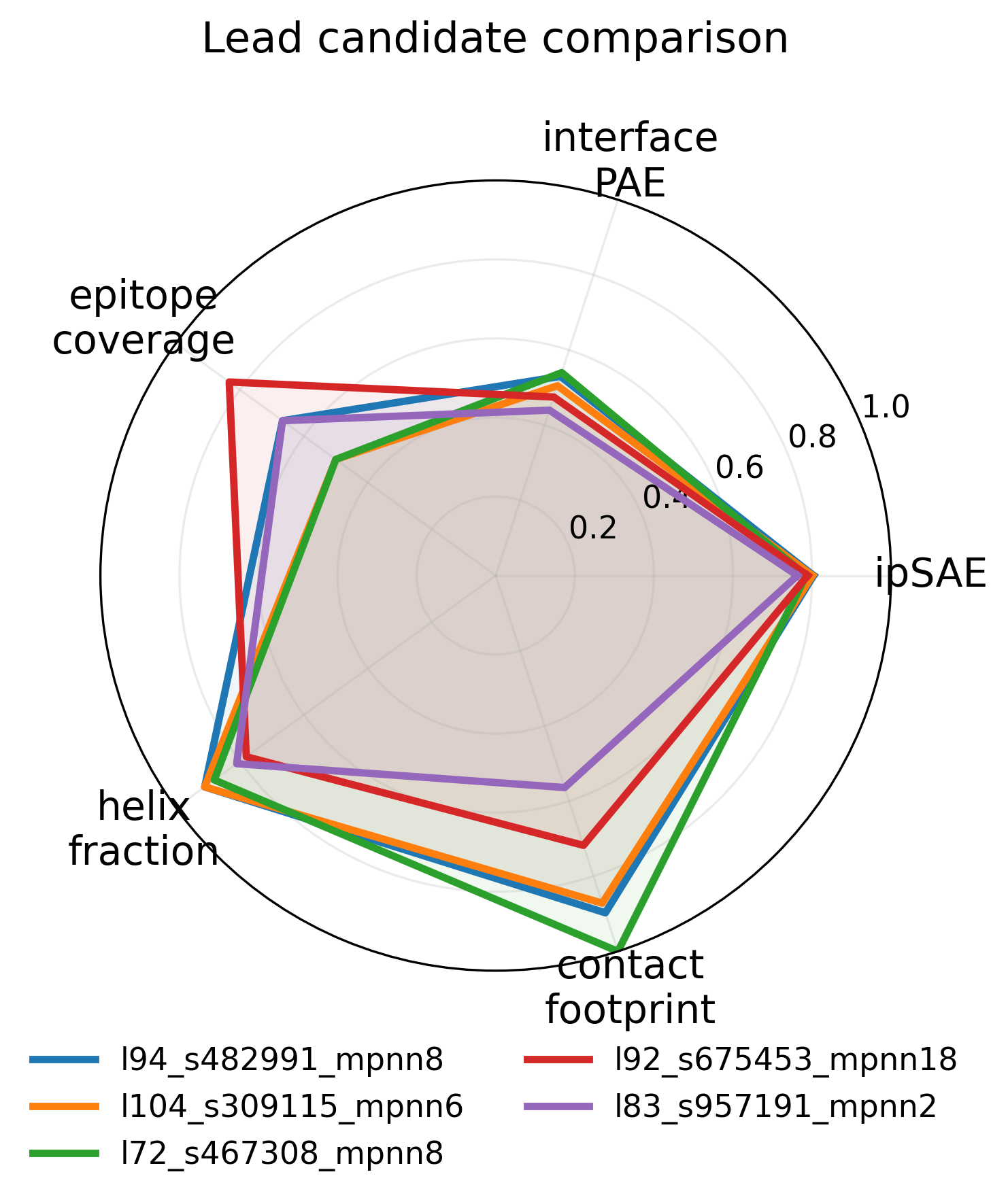}
    \caption{}
    \label{fig:leads-radar}
  \end{subfigure}
  \caption{Lead characterization. (a)~Epitope coverage: filled circles mark the hotspots
each lead engages (binder heavy atom within 5\,\AA{}, 3DUH chain-C numbering). No
single lead touches all six, but the five together cover every one. (b)~The five
leads across ipSAE, interface PAE, epitope coverage, helix fraction and contact
footprint, normalized to range; profiles are mutually distinct.}
  \label{fig:leads}
\end{figure}

We selected five leads by an explicit, non-purely-top-score rule: spanning both
target regimes, spanning the reliable ipSAE range rather than only its top, and
chosen so the set collectively covers all six hotspots, trading a little peak
ipSAE for regime and epitope coverage (Table~\ref{tab:leads}). All five are
compact all-$\alpha$-helical bundles of 72--104 residues. No single lead engages
all six hotspots, but the set collectively covers them (Fig.~\ref{fig:leads}a):
the central 80/81/89 patch by the highest-confidence p19-only leads, the
C-terminal 137/140 patch by l104, and the full-context lead l92 spanning five of
six (51/80/81/89/137). Per-hotspot coverage is C51~$\times4$, C80~$\times4$,
C81~$\times4$, C89~$\times4$, C137~$\times2$, C140~$\times1$, indicating the
hotspot set is addressable by multiple distinct predicted binders rather than
limited to one sub-patch.

ipSAE and Rosetta $\Delta$G (a relative estimate used as a filter, not a
calibrated binding free energy) are independent lines of evidence that agree on
overall quality while ranking the leads differently (Table~\ref{tab:leads}): l94
ranks first by ipSAE (0.804) yet has the smallest interface (2204\,\AA$^2$) and
fewest hydrogen bonds (3.5), whereas l72 ranks fourth by ipSAE (0.785) but is
strongest by $\Delta$G ($-87.3$ REU) with the largest interface. Reporting both
is therefore informative: they concur that all five are good and disagree on
which is best. The most physiologically credible lead is l92, which docks on p19
within the intact heterodimer and engages five of six hotspots, with the caveat
that its BindCraft hotspot RMSD (3.17\,\AA) exceeds that of the p19-only leads
($\sim$1.4--1.7\,\AA), indicating its docking geometry deviates somewhat from the
requested hotspot set even though it satisfies the 5-\AA{} contact criterion at
five of six positions.

\subsection{Structural characterization}
\begin{figure*}[!tbp]
  \centering
  \begin{subfigure}[t]{0.305\textwidth}
    \centering
    \includegraphics[width=0.864\textwidth]{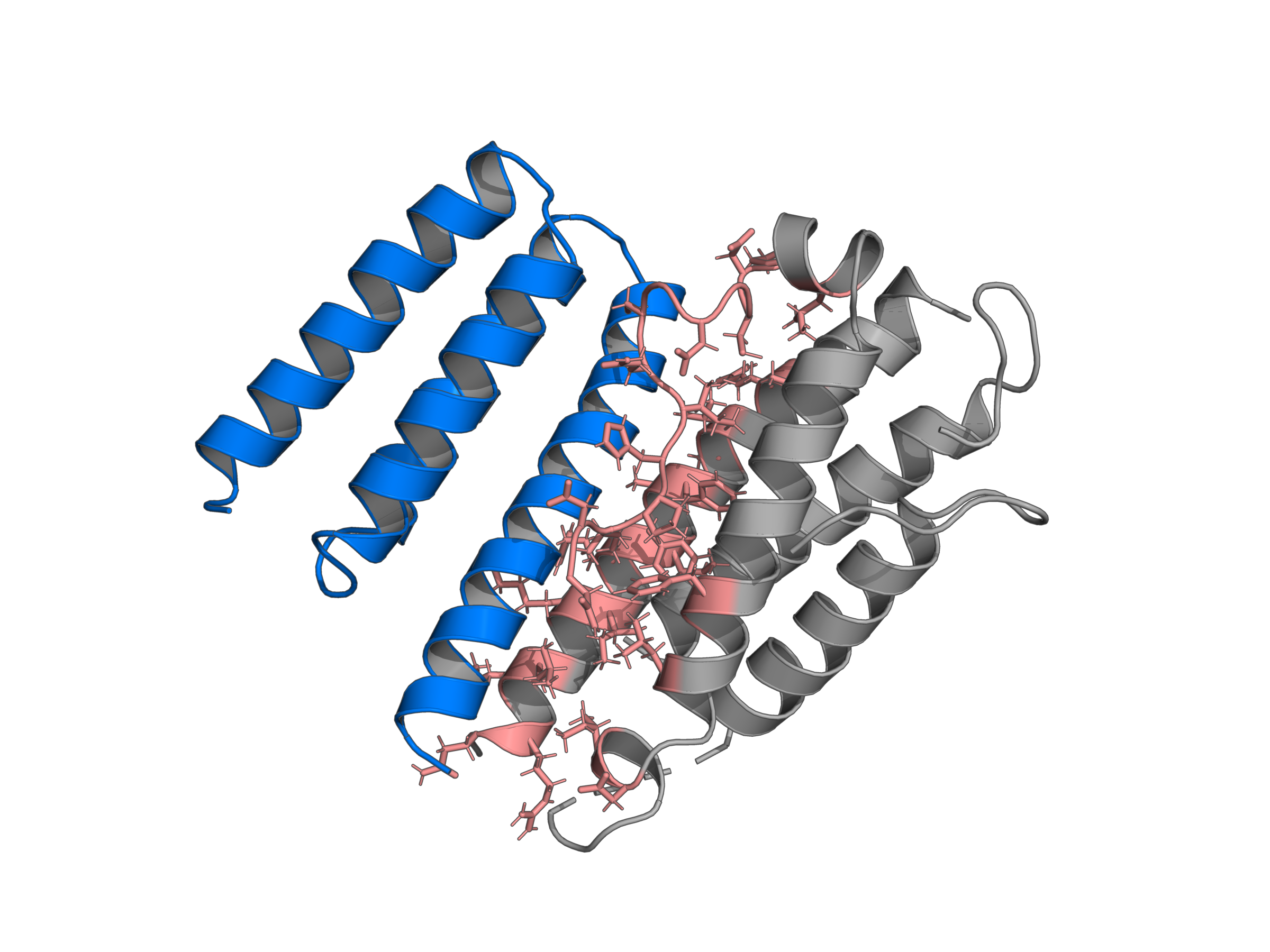}
    \caption{l94 (p19-only)}
  \end{subfigure}\hfill
  \begin{subfigure}[t]{0.305\textwidth}
    \centering
    \includegraphics[width=0.864\textwidth]{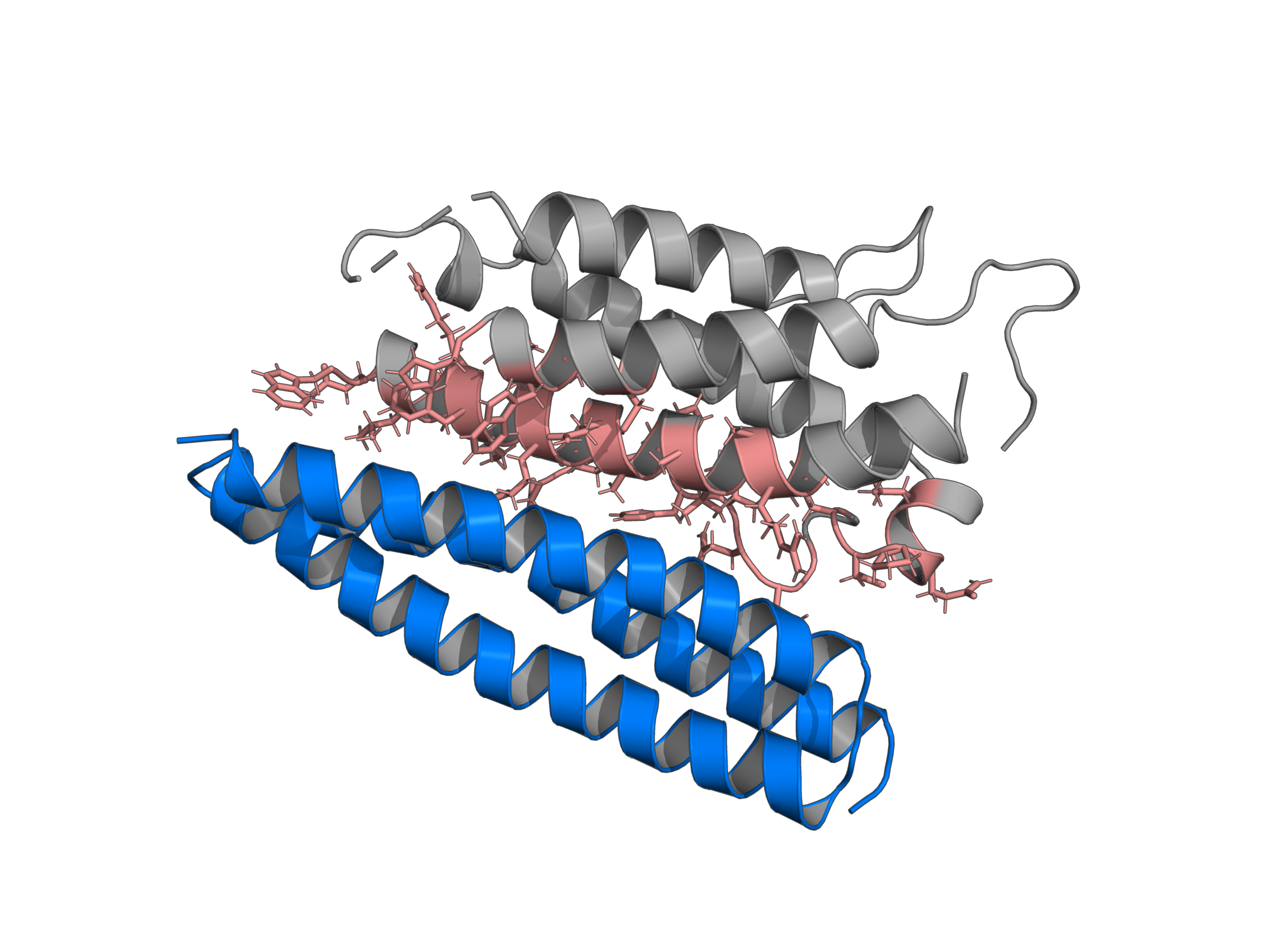}
    \caption{l104 (p19-only)}
  \end{subfigure}\hfill
  \begin{subfigure}[t]{0.305\textwidth}
    \centering
    \includegraphics[width=0.864\textwidth]{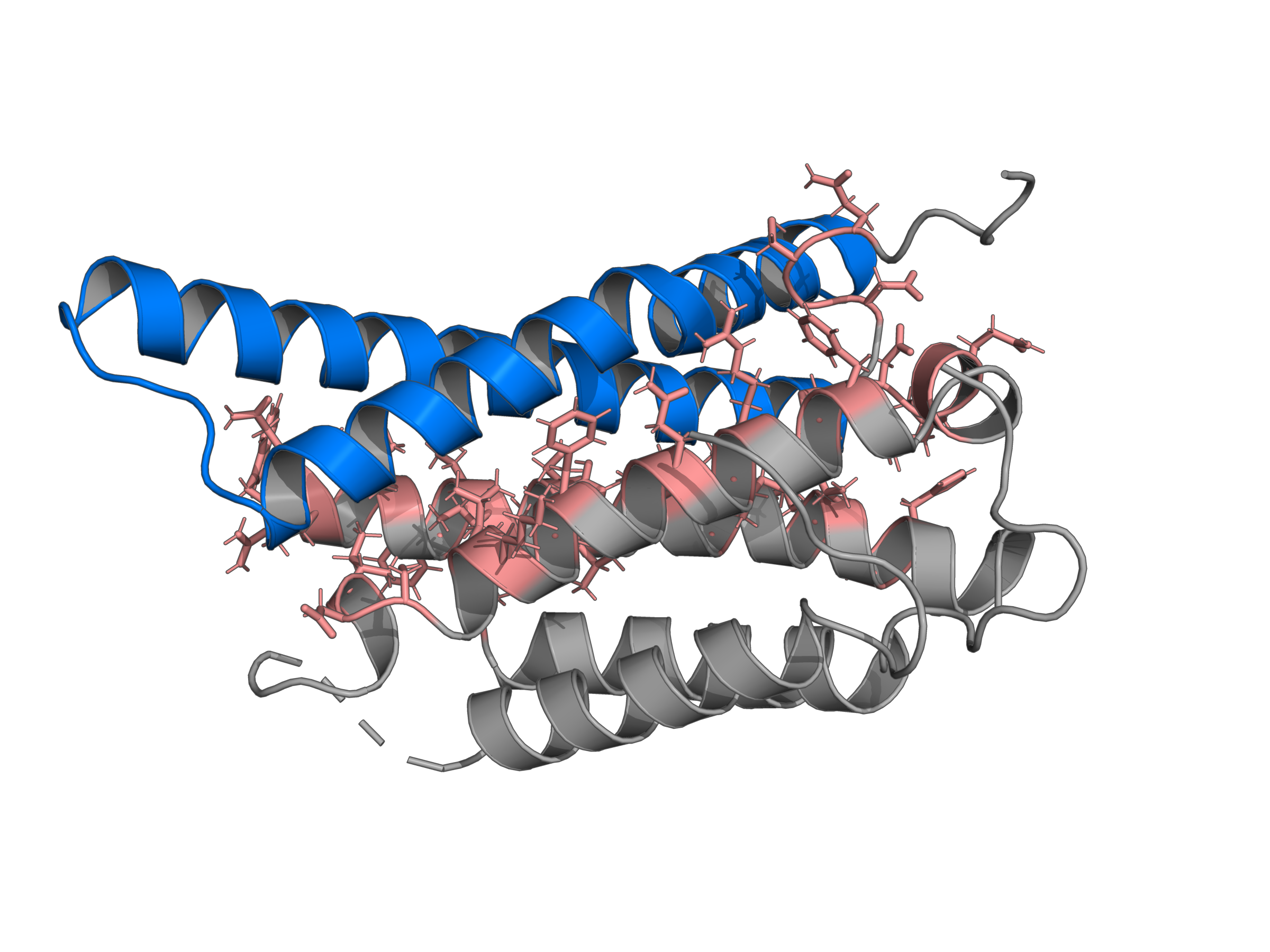}
    \caption{l72 (p19-only)}
  \end{subfigure}

  \vspace{5pt}

  \begin{subfigure}[t]{0.305\textwidth}
    \centering
    \includegraphics[width=0.864\textwidth]{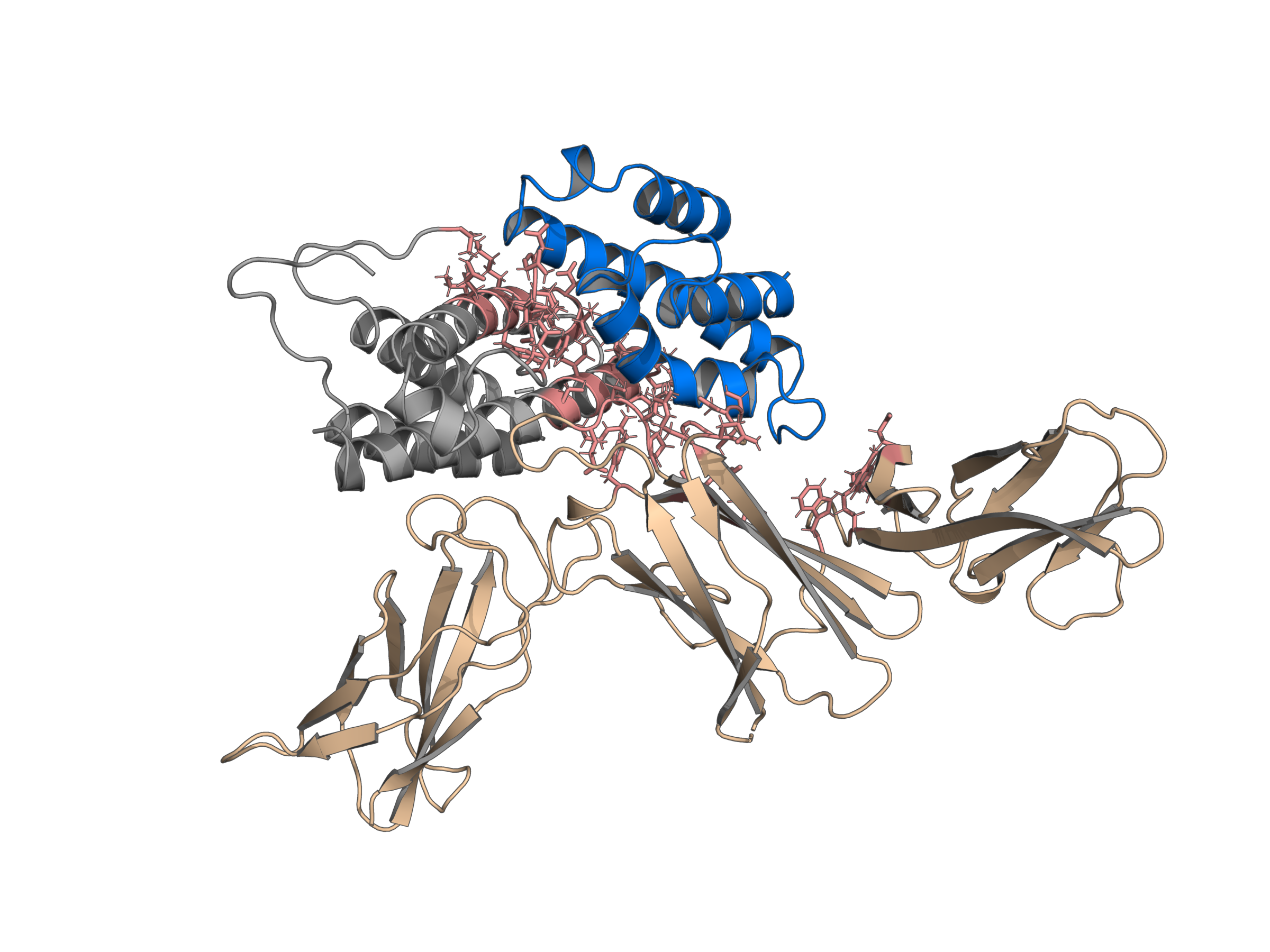}
    \caption{l92 (full heterodimer)}
  \end{subfigure}\hfill
  \begin{subfigure}[t]{0.305\textwidth}
    \centering
    \includegraphics[width=0.864\textwidth]{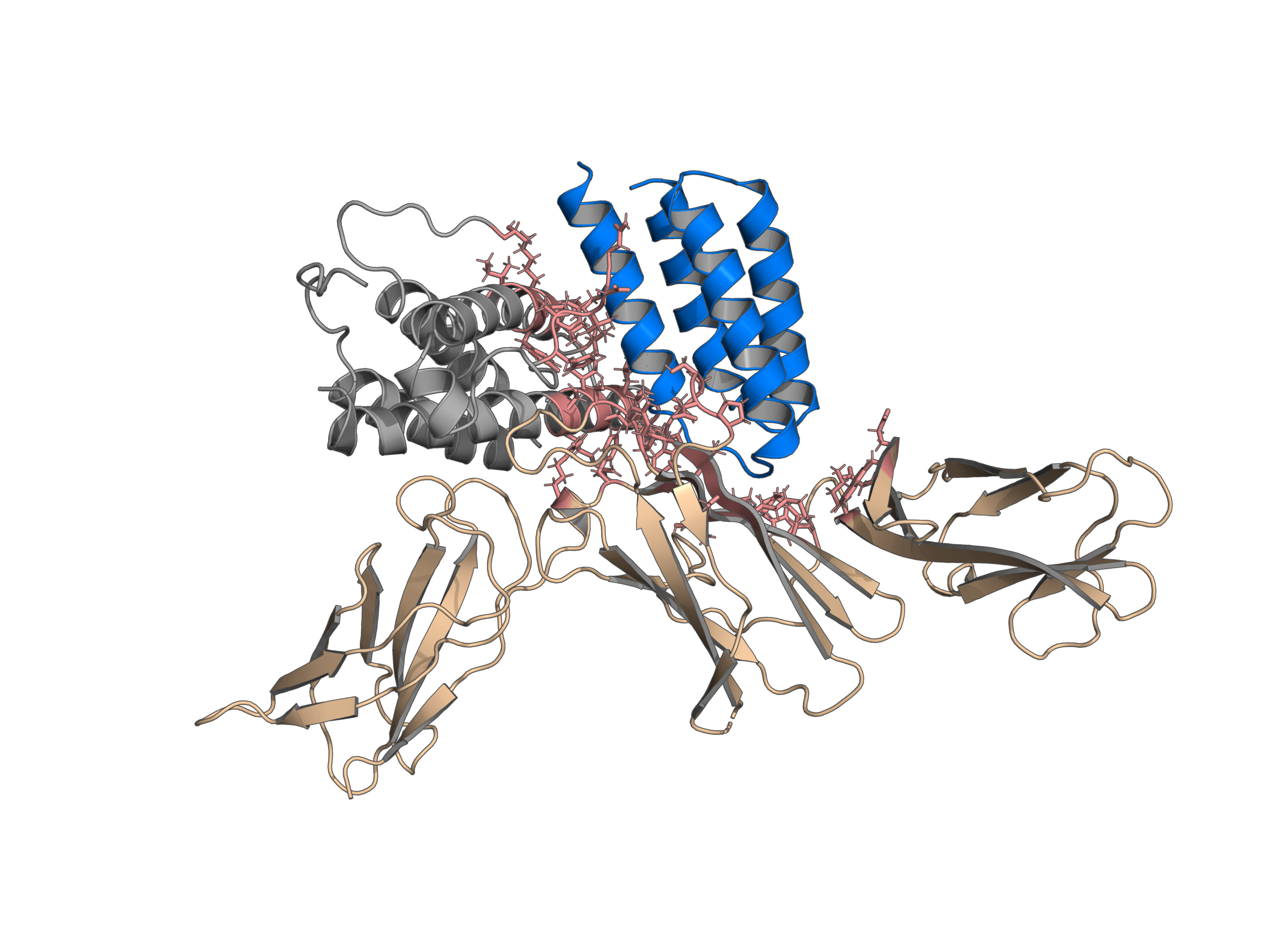}
    \caption{l83 (full heterodimer)}
  \end{subfigure}\hfill
  \begin{subfigure}[t]{0.305\textwidth}
    \centering
    \includegraphics[width=0.864\textwidth]{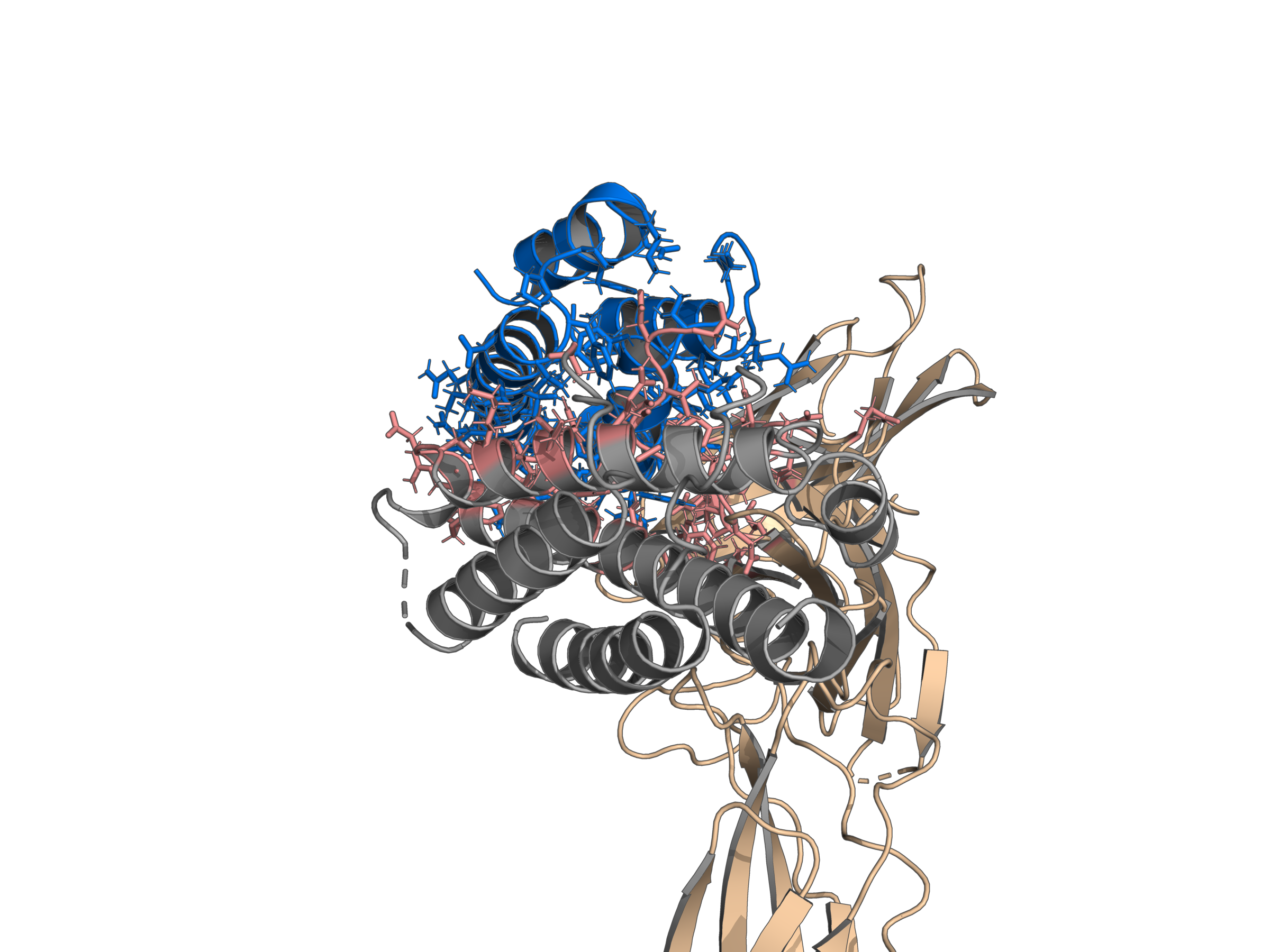}
    \caption{l92 interface close-up}
  \end{subfigure}

  \caption{Predicted structures of the five lead complexes (PyMOL). Each binder (blue) is
shown on p19 (gray) with interface residues within 5\,\AA{} as sticks; p40 (tan)
is shown for the two full-heterodimer leads. l92 docks on the p19 receptor face
even with p40 present, the configuration most relevant to neutralizing intact
IL-23.}
  \label{fig:renders}
\end{figure*}

All five leads are predicted $\alpha$-helical bundles (78--91\% helix, 0\%
$\beta$-strand by DSSP), ranging from a compact two-helix hairpin (l72, 72
residues) to three- and four/five-helix bundles, topologies and sizes favorable
for recombinant expression and chemical synthesis. Figure~\ref{fig:renders}
presents the predicted complexes.

\begin{table*}[!tbp]
  \centering
  \caption{Five lead candidates and their predicted properties. Every candidate clears the
$\geq$0.70 confident band while presenting a large buried interface and favorable
$\Delta$G, and the metrics disagree on the best lead (l94 by ipSAE, l72 by
$\Delta$G). $\Delta$G in REU; dSASA in \AA$^2$; SC $=$ shape complementarity;
HB $=$ interface hydrogen bonds; Cont. $=$ contacts within 5\,\AA; RMSD $=$
BindCraft hotspot RMSD. Sequences in Supplementary~S1.$^{\dagger}$}
  \label{tab:leads}
  \footnotesize
  \setlength{\tabcolsep}{3.5pt}
  \begin{tabular}{cllrrrrlrrrrrr}
    \toprule
    \# & Design & Regime & Len & ipSAE & minPAE (\AA) & RMSD (\AA) & Hotspots (n/6) & \%Hel
       & $\Delta$G (REU) & dSASA & SC & HB & Cont. \\
    \midrule
    1 & l94\_s482991\_mpnn8  & p19-only & 94  & 0.804 & 0.94 & 1.69 & 51,80,81,89 (4)     & 91 & $-57.5$          & 2204 & 0.68 & 3.5  & 35 \\
    2 & l104\_s309115\_mpnn6 & p19-only & 104 & 0.801          & 0.99 & 1.50 & 51,137,140 (3)      & 91 & $-84.9$          & 2962 & 0.69 & 10.5 & 34 \\
    3 & l72\_s467308\_mpnn8  & p19-only & 72  & 0.785          & 0.92 & 1.37 & 80,81,89 (3)        & 88 & -87.3 & 3018 & 0.68 & 8.5  & 39 \\
    4 & l92\_s675453\_mpnn18 & full     & 92  & 0.789          & 1.05 & 3.17 & 51,80,81,89,137 (5) & 78 & $-77.0$          & 2960 & 0.62 & 9.5  & 28 \\
    5 & l83\_s957191\_mpnn2  & full     & 83  & 0.763          & 1.12 & 1.86 & 51,80,81,89 (4)     & 81 & $-75.0$          & 2943 & 0.67 & 9.0  & 22 \\
    \bottomrule
  \end{tabular}

  \smallskip
  {\footnotesize $^{\dagger}$Designs are the five leads of \S\,IV-D; sequences in Supplementary~S1; full metric table in Supplementary~S4.
  $\Delta$G, dSASA, SC, and HB are BindCraft
  per-design Rosetta metrics: the two full-heterodimer leads (l92, l83) from the
  full-heterodimer design statistics and the three p19-only leads (l94, l104,
  l72) from the p19-only design statistics.}
\end{table*}

\subsection{No predicted p40 interface}
Because p40 is shared beyond IL-23, p40-directed off-target binding is a key risk
for any IL-23 therapeutic, so we frame the selectivity result narrowly. Three
observations support p40 sparing. All six design hotspots lie on the p19 receptor
face, not on p40. When the conventional designs were re-scored against p40 the
best p40 interface reached only ipSAE 0.147 (versus 0.235 on p19). Most directly,
we ran a p40-alone counter-screen of all five leads, co-folding each binder with
p40 in ColabFold (\texttt{alphafold2\_multimer\_v3}) and scoring with the same
ipSAE metric (Table~\ref{tab:p40}); ColabFold-Multimer rather than initial-guess
was used here to avoid forcing the IL-23 pose onto an off-target protein. Every
lead returned ipSAE 0.00 against p40 (ipTM 0.11--0.12) versus 0.76--0.80 on p19,
and the null held under both a fast single-model screen and a full five-model,
three-recycle re-run. Notably l92 and l83 were designed in the presence of p40
and still register 0.00, so this is not merely a consequence of never having seen
the subunit. We note what this assay does and does not establish: it tests
whether a binder forms a spurious interface with p40 as a separate molecule, and
the answer is uniformly negative; it does not test whether a designed pose on p19
is geometrically compatible with p40 in the intact cytokine, where the subunits
are covalently linked. That assembly-level question requires superposition into
an experimental IL-23 structure rather than a two-chain prediction and is the
subject of companion work in preparation. We therefore restrict the claim to the
absence of a detectable p40 interface.

\begin{table}[!tbp]
  \centering
  \caption{Counter-screen of the five leads against p40 alone. The p40 ipSAE comes from
co-folding each binder with p40 alone and scoring with the identical metric.
Every lead collapses to 0.00, far below its p19 value and the 0.60 threshold.
Raw scores: Supplementary~S3.}
  \label{tab:p40}
  \small
  \setlength{\tabcolsep}{10pt}
  \renewcommand{\arraystretch}{1.3}
  \begin{tabular}{lccc}
    \toprule
    Lead & ipSAE (p19) & ipSAE (p40) & ipTM (p40) \\
    \midrule
    l94  & 0.804 & 0.00 & 0.12 \\
    l104 & 0.801 & 0.00 & 0.12 \\
    l72  & 0.785 & 0.00 & 0.11 \\
    l92  & 0.789 & 0.00 & 0.12 \\
    l83  & 0.763 & 0.00 & 0.12 \\
    \bottomrule
  \end{tabular}
\end{table}

\section{Discussion}
The central finding is methodological as much as biological. Two
structure-conditioned pipelines, evaluated by an identical interface-aware
scorer, gave opposite outcomes on the same target: the conventional
fixed-backbone pipeline produced confident folds but no reliable interface
(0/972), whereas interface-optimized hallucination produced reliable interfaces
at high rate (42/50). One caveat, noted in Methods~E, is that the scorer is not
fully independent of BindCraft's objective; two observations argue against this
being the explanation, since the gap persists under per-attempt normalization on
comparable budgets and reproduces under interface geometry alone
(Table~\ref{tab:geom}). The difference is that BindCraft co-optimizes binder and
interface and permits induced fit, while the conventional pipeline commits to a
backbone before the interface is evaluated. On a shallow, helical
receptor-binding surface like p19's IL-23R epitope, that distinction appears
decisive.

The scoring artifact (Section~IV-A) is a cautionary point of independent
interest: trusting single-sequence ColabFold would have wrongly condemned the
binders as unfoldable, so initial-guess-seeded scoring matters not only for
ranking but for avoiding false negatives. The leads make the same point from the
other direction: BindCraft's design-time ipTM spans only 0.86--0.92 across the
five, a near-saturated band assigning almost identical confidence to designs
whose ipSAE and $\Delta$G differ substantially and that disagree on the best
candidate (l94 by ipSAE, l92 by ipTM, l72 by $\Delta$G). Combined with its
uninformative behaviour on the conventional failures, this shows ipTM losing
discriminative power at both ends of the quality scale.

Biologically, the five leads collectively cover the targeted IL-23R face, and the
full-heterodimer lead l92 is the most physiologically relevant since it engages
p19 in the presence of p40, though its higher hotspot RMSD warrants scrutiny. Of
the 24 p19 residues forming the IL-23R epitope, guselkumab overlaps 58\% and
risankizumab only 8\%, the remainder of their footprints lying off the receptor
site~\cite{daniele2024epitope}. Our binders were conditioned on the
IL-23R-binding face, so they most closely recapitulate guselkumab's
receptor-competitive mode, and the six-hotspot epitope could be shifted to tune
between competition and steric blockade. The \emph{neutralizer} designation rests
on where and how much the binders bury, not on a functional assay: the hotspots
were selected as IL-23R contact residues, so a binder burying 2204--3018~\AA$^2$
over that face (Table~\ref{tab:leads}) occludes the receptor footprint
sterically, making steric competition the most parsimonious predicted mechanism
until a competition or STAT3 assay is performed.

\textbf{Limitations.} All results are computational: ipSAE is a predictor, not a
guarantee, and designed binders can still fail experimentally. The target is a
single, partially modeled crystal conformation (p19 has unmodeled loops at
30--46, 95--99, 126--136), and we did not measure affinity. Three further
limitations bound the comparison. The conventional pipeline was run
score-after-design, with AlphaFold-initial-guess applied once to finished designs
rather than inside an iterative generate--filter--regenerate loop, so the
negative result characterizes this pipeline at the budget of Methods~E, not the
method's ceiling. The 0.60 and 0.70 bands are AlphaFold-DB homodimer ranges, not
calibrated for de novo binder design, and a score at the floor cannot distinguish
genuine non-interaction from a prediction in which the chains never contact.
Finally, our selectivity test asks only whether a binder forms a spurious
interface with isolated p40; steric compatibility with p40 inside the intact
covalently linked cytokine is a separate assembly-level question we did not
evaluate.

\textbf{Future work.} Computationally, the next steps are an orthogonal
fold/interface predictor (AlphaFold3, Boltz-2 or Chai-1) as an independent check
on the leads, and assembly-level steric validation against an experimental IL-23
structure. Experimentally, the next steps are recombinant expression, affinity
measurement by biolayer interferometry or surface plasmon resonance, a cell-based
STAT3 blockade assay to test neutralization, and ideally a co-structure to
confirm the predicted epitope. The leads are intentionally diverse, so one topology failing still leaves independent starting points.

\section{Conclusion}
We presented a uniformly scored comparison of de novo binder-design pipelines
against the IL-23R-binding face of the IL-23 subunit p19. A conventional pipeline
produced no reliable predicted interface across 972 designs under interface-aware
scoring, after we first showed the discouraging single-sequence ColabFold signal
was a scoring artifact; the same failure reproduces under AlphaFold-independent
interface geometry. An interface-optimized pipeline, scored identically, produced
42 reliable designs of 50 (38 confident), from which we characterized five
all-$\alpha$ leads collectively engaging all six targeted hotspots, two within
the intact heterodimer. These appear to be the first computationally proposed de
novo miniprotein candidates designed to occlude the IL-23R-binding face of the
IL-23 cytokine directly, providing a starting point for experimental validation.

\section*{Acknowledgment}
The authors acknowledge the University of Maryland's Zaratan high-performance
computing cluster for the computational resources used in this work.

\end{document}